\documentclass[12pt,twoside]{article}   %For printing on two sides of page
\usepackage[section]{placeins}   %
\usepackage{graphicx}

\newcommand{\cen}[1]{\begin{center} #1 \end{center}}

\usepackage[mathlines]{lineno}
\usepackage{hyperref}
\hypersetup{ colorlinks,
    citecolor=blue,
    filecolor=blue,
    linkcolor=blue,
    urlcolor=blue
}

\usepackage{xcolor}
\definecolor{gray}{rgb}{0.6,0.6,0.6}
\definecolor{red}{rgb}{0.85,0,0}
\definecolor{green}{rgb}{0,0.85,0}
\definecolor{blue}{rgb}{0,0,0.85}
\definecolor{beige}{rgb}{0.92,0.87,0.78}
\usepackage[all]{hypcap}    %causes link to figures to go to figure, not caption
\usepackage{array,booktabs,multirow,authblk}
\newcommand{\rom}[1]{\uppercase\expandafter{\romannumeral #1\relax}}
\newcolumntype{C}[1]{>{\centering\let\newline\\\arraybackslash\hspace{0pt}}m{#1}}
\usepackage{tikz}
\usetikzlibrary{calc}

\begin{document}

\cen{\sf {\Large {\bfseries RESECT-SEG: Open access annotations of intra-operative brain tumor ultrasound images } \\
\vspace*{10mm}
Bahareh Behboodi\textsuperscript{1,2,*},
Francois-Xavier Carton\textsuperscript{3,*}
Matthieu Chabanas\textsuperscript{3},
Sandrine De Ribaupierre\textsuperscript{4},
Ole Solheim\textsuperscript{5,6},
Bodil K. R. Munkvold\textsuperscript{5,6},
Hassan Rivaz\textsuperscript{1,2},
Yiming Xiao\textsuperscript{1,2,},
Ingerid Reinertsen\textsuperscript{7,8,**}}\\
\textsuperscript{1}Department of Electrical and Computer Engineering, Concordia University, Montreal, Canada\\
\textsuperscript{2}PERFORM Centre, Concordia University, Montreal, Canada \\
\textsuperscript{3}Université Grenoble Alpes, CNRS, Grenoble INP, TIMC, F-38000 Grenoble, France\\
\textsuperscript{4}Department of Clinical Neurological Sciences, Schulich School of Medicine and Dentistry, Western University, London, Ontario, Canada\\
\textsuperscript{5}Department of Neurosurgery, St. Olavs Hospital, Trondheim University Hospital, Norway\\
\textsuperscript{6}Department of Neuromedicine and Movement Science, Norwegian University of Science and Technology (NTNU), Trondheim, Norway\\
\textsuperscript{7}Department of Health Research, SINTEF Digital, Trondheim, Norway\\
\textsuperscript{8}Department of Circulation and Medical Imaging, Norwegian University of Science and Technology (NTNU), Trondheim, Norway
\vspace{5mm}\\
%Version typeset \today\\
\textsuperscript{*} These authors share first authorship\\
\textsuperscript{**} Corresponding author: Ingerid.Reinertsen@sintef.no\\
}

\pagenumbering{roman}
\setcounter{page}{1}
\pagestyle{plain}
%Correspondence: Ingerid Reinertsen. email: Ingerid.Reinertsen@sintef.no
%*\\
%$^{\star}$ These authors share the first authorship.
% note, probably best not to use a student's e-mail as it won't be valid for
% very long.

\begin{abstract}
\noindent {\bf Purpose:}
Registration and segmentation of magnetic resonance  (MR) and ultrasound (US) images  play an essential role in surgical planning and resection of brain tumors. However, validating these techniques is challenging due to the scarcity of publicly accessible sources with high-quality ground truth information. To this end, we propose a unique annotation dataset of tumor tissues and resection cavities from the previously published RESECT dataset \cite{xiao2017re} to encourage a more rigorous assessments of image processing techniques.\\
{\bf Acquisition and validation methods:}
The RESECT database~\cite{xiao2017re} consists of MR and intraoperative US (iUS) images of 23 patients who underwent resection surgeries. The proposed dataset contains tumor tissues and resection cavity annotations of the iUS images. The quality of annotations were validated by two highly experienced neurosurgeons through several assessment criteria. \\
{\bf Data format and availability:} Annotations of tumor tissues and resection cavities are provided in 3D NIFTI formats. Both sets of annotations are accessible online in the \url{https://osf.io/6y4db}.   \\
{\bf Discussion and potential applications:} The proposed database includes tumor tissue and resection cavity annotations from real-world clinical ultrasound brain images to evaluate segmentation and registration methods. These labels could also be used to train deep learning approaches. Eventually, this dataset should further improve the quality of image guidance in neurosurgery.\\

\end{abstract}

% \newpage     %may or may not be needed

% The table of contents is for drafting and refereeing purposes only. Note
% that all links to references, tables and figures can be clicked on and
% returned to calling point using cmd[ on a Mac using Preview or some
% equivalent on PCs (see View - go to on whatever reader).
% \tableofcontents

% \newpage

\setlength{\baselineskip}{0.7cm}      %double spacing

\pagenumbering{arabic}
\setcounter{page}{1}
%\pagestyle{fancy}

% #################################################### Purpose
\section{Purpose}
Gliomas are the most common malignant primary brain tumors in the central nervous system (CNS) and originate from glial cells \cite{omuro2013glioblastoma}. According to the World Health Organisation (WHO) scheme, gliomas are classified and graded into grade \rom{1}-\rom{4} \cite{louis2021}, where grades \rom{1}-\rom{2} and  \rom{3}-\rom{4} are considered as low- and high-grade tumors, respectively. Despite recent advances, the prognosis remain dismal for the most common and aggressive types and treatment options are still limited. Maximal safe surgical resection of the tumor remains an essential part of the standard treatment for both low- and high-grade tumors  \cite{schneider2010gliomas}. In neurosurgery, image based navigation systems are widely used for planning and guidance.

Before the surgery, MR imaging is used to characterize and identify the exact location of the tumor. However, during surgery the brain tissue will deform due to edema and gravity (i.e. brain shift). Thus, the images need to be updated to maintain the accuracy of the surgical navigation. The use of intraoperative imaging techniques such as MRI and ultrasound (US) can provide an updated road map to accurately guide the surgeon to reach the exact location of the tumor \cite{nabavi2001serial,reinertsen2021essential}. Acquiring data during different stages of the surgery helps surgeon to better monitor the progress of resection, and consequently, operate more precisely. Interaoperative MRI (iMRI) offers images with higher spatial resolution and dynamic range in comparison with interaoperative US (iUS) \cite{bastos2021challenges}. Although iMRI is considered the gold standard interaoperative imaging modality for brain tumor surgery, it is costly, adds a long time to the operation and requires dedicated operating rooms \cite{yeole2020navigated,zhang2015impact}. In contrast, iUS has emerged as a portable, low cost, flexible, and versatile modality that presents real-time scanning without altering the surgical workflow \cite{sastry2017applications,bastos2021clinical}. Due to the easy procedure of acquiring iUS rather than iMRI, several recent studies have shown impacts of iUS during the resection surgeries \cite{vstevno2021current,rubin1980intraoperative,sastry2017applications, yeole2020navigated}.

While iUS present a number of advantages in the context of brain tumor resection, US images can be difficult to interpret. Non-standard imaging planes and unfamiliar contrast are major factor limiting the efficient use of ultrasound in neurosurgery. Automatic segmentation of important structures such as the tumor in the images could ease the interpretation and enable more widespread use of this modality.

Brain tumor's margin, shape, and its location can be identified through image segmentation that is considered as a crucial step towards accurate assessment of successful resection. Many studies have explored the benefits of recently developed deep learning-based algorithms in brain tumor segmentation in both US \cite{v2019learning,carton2020automatic,NAMBURETE20181,9453850,ilunga2018patient,canalini2019segmentation} and MRI \cite{canalini2020enhanced,DBLP:journals/corr/MilletariAKPRML16, dong2017automatic,li2019overfitting,spitzer2018improving}. Because of the BRATS challenge, which provided large data with annotations, there have been  considerable advancements in brain tumour segmentation from preoperative MRI images~\cite{menze2014multimodal,bakas2018identifying}. Despite the great progress that these algorithms have provided in finding solutions to various medical segmentation problems, their performance is highly dependent on the availability of large, high-quality expert-annotated datasets. Furthermore, another challenge faced by researchers especially in the field of medical imaging is finding real clinical images with which to validate new image analysis algorithms \cite{tajbakhsh2020embracing}. Preparing both data and annotations of medical images, especially iUS images, are expensive and tedious to acquire which is the main reason that such datasets are rarely available to public.

To date, there are only two main datasets that provide intraoperative brain images. The first database is the BITE dataset presented by Montreal Neurological Institute (MNI) in 2012 \cite{mercier2012online}. It contains preoperative and postoperative MRI scans as well as multiple iUS images of 14 patients. The second  is the RESECT database that consists of preoperative MRI scans, and iUS images from 23 patients with low-grade gliomas \cite{xiao2017re} released in 2017. None of the abovementioned datasets provide annotations of their iUS images. Tumors in RESECT iUS images were annotated by Munkvold et al.~\cite{Munkvold2018} for 19 case, although these labels were not publicly available. Resection cavities were first annotated by Carton et al.~\cite{carton2020} and then Canalini et al.~\cite{canalini2020enhanced}. This later group made their annotations public, for a limited number of cases and with limited validation.

In this work, we present expert-annotations of brain tumor and resection cavities in iUS images of RESECT database. The aim of our proposed iUS annotations is to help the intraoperative guidance for better resection outcomes. In the iUS 3D volumes taken before the resection starts, we focus on an optimal delineation of the tumor. After resection we focus on clearly identifying the resection cavity, which can be an important factor both to evaluate the resection progress and to enhance the registration of images acquired before and all along surgery. In the following sections we detail the expert-annotation protocols for segmenting tumor and resection cavities for all 23 iUS 3D scans, which were further validated by two experienced neurosurgeons.

% #################################################### Acquisition and Validation Methods
\section{Acquisition and Validation Methods:}
\subsection{RESECT Database}
The RESECT database consists of preoperative contrast enhanced T1-weighted and T2 FLAIR MRI scans alongside three 3D volumes of iUS scans for 23 clinical patients with low-grade gliomas (grade \rom{2}) who underwent surgeries between 2011 and 2016 at St. Olavs University Hospital, Trondheim, Norway. The image acquisition protocols are described in detail in \cite{xiao2017re}. A summary is provided here for completeness. The iUS scans were taken at 3 different timepoints during the procedure: (1) The first volume acquired on the intact dura or the cortical surface before the resection starts, (2) the second one acquired during the resection, and (3) the last one was collected after the resection for controlling purposes. All the ultrasound images in RESECT database were acquired by expert surgeon.  The details of acquisition procedure can be summarized as follow:
\begin{itemize}
    \item Preoperative MRI scans: T1-weighted and T2 FLAIR sequences acquired on 3T Magnetom Skyra MRI scanner both with 1 mm isotropic voxel size, except three patients who underwent the MR imaging on a 1.5T Magnetom Avanto MRI scanner with 1 mm slice thickness.
    \item Intraoperative US scans: 3D US images collected using 12FLA-L linear probe of Sonowand Invite neuronavigation system with a frequency range of 6-12 MHz while integrated to the NDI Polaris optical tracking system.
\end{itemize}

The RESECT database also contains landmarks manually selected by experts in neuroanatomy. These pairs of homologous points allow registering the MRI to the corresponding US volume. The registered MRI volumes have a better alignment with the iUS volumes, and were used as a guide for segmentation brain structures. %MC: and +falx, sulci, ventricles of course

\subsection{iUS Tumor Annotation Protocol}
We used annotations of MR images to initialize the ROI for the tumor in iUS images. Therefore, here we first explain the details of MR delineation which followed the protocol explained in \cite{stensjoen2015growth}. We used the NIFTI formats of MRI tumor annotations from the previous study presented by Stensj{\o}en et al.~\cite{stensjoen2015growth}, as the initial iUS annotations. In our iUS annotation protocol, the ground truth annotations were performed on the acquired iUS images using 3D Slicer \cite{fedorov20123d}. Due to the brain-shift, the tumor boundaries in US and MRI images were not the same~\cite{xiao2017re}. Therefore, as the US and MRI images are usually interpreted together during the resection surgery, MR tumor annotations needed to first be registered to the iUS images using the available landmarks in the RESECT database. This step helped to compensate the misalignment of tumor border in iUS and MR images. Consequently, the MR annotations were loaded to the Slicer scene and used as the initial iUS delineation. In addition, we used 'Label Map Smoothing' from 'Surface Models', one of the existing modules in the Slicer,  to smooth the surface of 3D tumor segment. In this regard, in the antialiasing parameters setup, we initialized the number of iterations to 150 with the maximum root mean square (RMS) error of 0.1 to avoid excessive smoothing. Finally, the smoothed tumor segment was further corrected manually to ensure the segmentation mask correctly covers the tumor region in iUS image. In this procedure, we additionally took advantage of FLAIR MR images, registered to iUS images to avoid brain-shift effects, as the complementary overlay guidance to generate finer iUS tumor annotations.

\subsection{iUS Resection Cavity Annotation Protocol}
The resection cavity is generally surrounded by hyperechoic signal in iUS images (figure~\ref{fig:resec-example}), which can be due to several factors.
A first one is the difference of sound attenuation between tissue and the saline water used to fill the cavity, which creates an hyperechoic artifact. Blood remaining in the cavity or on its borders is also more hyperechoic than the saline water solution.
These artifacts impair the estimation of the boudaries of the resection cavity, especially at the bottom of the cavity.
Thus, we decided to exclude this hyperechoic signal surrounding the cavities from the segmentation, to avoid false positives as much as possible. Only the clear, darker signal was considered as cavity.
This can lead to a slight underestimation of some cavities, especially as blood may have been excluded from the segmentation while it is technically part of the cavity.
In three extreme cases (Case 11 during and after resection, and Case 15 during resection) the cavity was very small and totally filled with blood, with no dark signal inside. Since the cavity border could not be differentiated from the hyperechoic signal, no voxels were labeled as cavity for these cases.

\definecolor{resecblue}{RGB}{0,191,255}

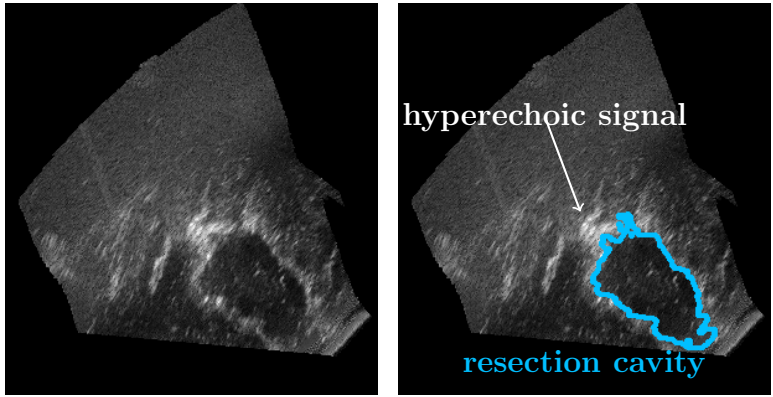
\begin{figure}
    \centering
    \begin{minipage}{0.3\textwidth}
        \includegraphics[width=\textwidth]{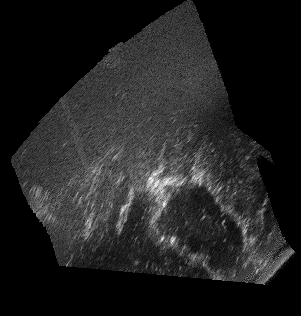}
    \end{minipage}
    \begin{minipage}{0.3\textwidth}
        \begin{tikzpicture}
            \node (p) {\includegraphics[width=\textwidth]{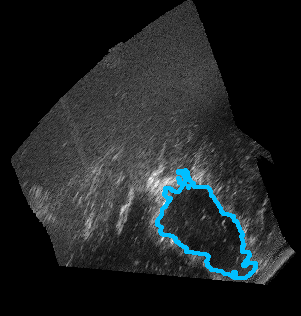}};
            \coordinate (o) at (p.south west);
            \coordinate (w) at ($(p.east) - (p.west)$);
            \coordinate (h) at ($(p.north) - (p.south)$);
            \draw[<-, thick, white, font=\bf] ($(o) + 0.49*(w) + 0.47*(h)$) -- ($(o) + 0.4*(w) + 0.7*(h)$) node {hyperechoic signal};
            \node[color=resecblue, font=\bf] at ($(o) + 0.5*(w) + 0.1*(h)$) {resection cavity};
        \end{tikzpicture}
    \end{minipage}
    \caption{Ultrasound image of a resection cavity.}
    \label{fig:resec-example}
\end{figure}

The resection cavities were manually segmented using a proprietary software. A set of regularly spaced slices, about one every five slices, was manually delineated.
The remaining slices were then filled using ITK's morphological interpolation with the c3d command-line tool, part of itksnap~\cite{py06nimg}.

Most of the RESECT cases were initially segmented by two raters for a previous study~\cite{carton2020}. After reporting intra- and inter- rater variability, these segmentations were reviewed by a neurosurgeon and edited accordingly. For this study, the remaining RESECT cases were first segmented, then all segmentations were refined during the validation protocol described in the following section.

\subsection{Data Validation}
\begin{figure}
  \centering
  \centerline{\includegraphics[width=16cm]{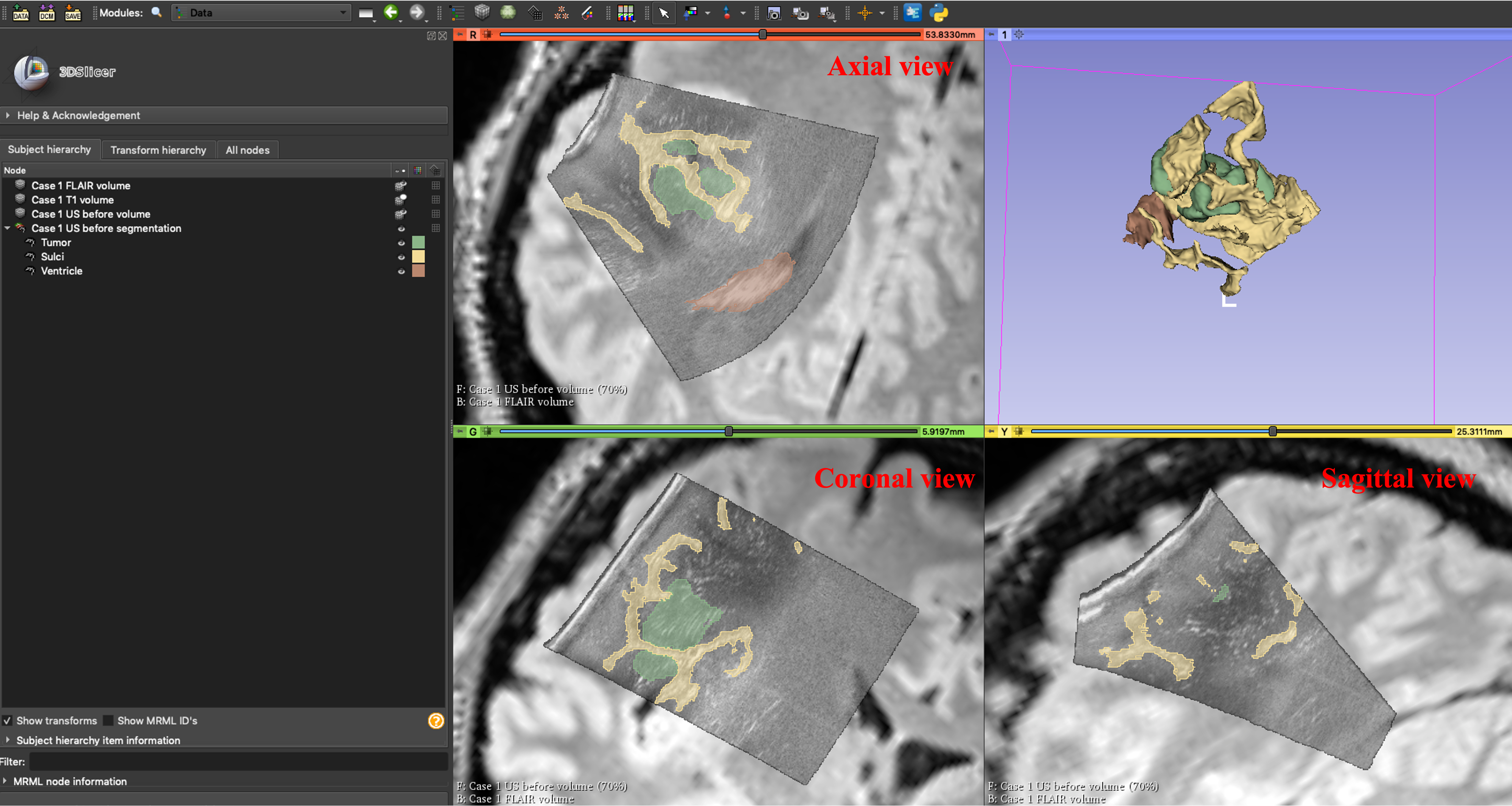}}
  \caption{An example view of iUS 3D scan overlaid with tumor annotations (MR image is as reference guide) in 3D Slicer. (Green: tumor, yellow: sulci, brown: ventricle)}
  \label{fig:tumor-sulci}
\end{figure}

Both tumor and resection cavity annotations presented in this work were validated by two experienced specialists. The annotations were presented to the specialists through a 3D Slicer scene including the original iUS image, segmentation masks as well as FLAIR MR images, as the reference. Figure~\ref{fig:tumor-sulci} represents an example of such scene. We asked specialists to grade both tumor and resection cavity segmentation masks based on four main criteria:
\begin{itemize}
    \item Smoothness of the tumor boundaries (both tumor and cavities) (SMT)
    \item Identification of cancerous tissues (IdT)
    \item Exclusion of non-cancerous tissues (ExT)
    \item Quality of resection cavity (IdR)
\end{itemize}

The grading scheme for each criteria was in the scale of 1 to 5 defined as major improvement needed, minor improvement needed, acceptable, good quality, and excellent, respectively. For each criteria, an average score of 3 was needed to pass the quality control of annotation masks, otherwise the masks were revised according to the requirements.

% #################################################### Data format and usage notes
\section{Data Format and Availability}
\subsection{Data Format}
The preoperative T1-weighted and T2 FLAIR MR images as well as iUS images of the RESECT database were distributed in MINC2 and NIFTI formats. Therefore, the proposed tumor and resection cavity annotations of iUS images were also deposited in NIFTI formats. Annotations of tumor and cavity were generated separately and can be viewed in many visualization software such as 3D Slicer. Moreover, the NIFTI format is compatible for use in deep learning-based packages. In our proposed annotation dataset, each patient is assigned to a folder containing its corresponding tumor and cavity annotations. We kept the folder IDs of patients the same as RESECT database for simplicity. For example, considering case 1 in RESECT database, the corresponding three iUS scans were named in the form of 'Case1-US-before.nii.gz', 'Case1-US-during.nii.gz', and 'Case1-US-after.nii.gz'. Similarly, in our proposed annotations dataset, the corresponding tumor and resection cavity annotation masks for case 1 were as 'Case1-US-before-tumor.nii.gz', 'Case1-US-during-resection.nii.gz', 'Case1-US-after-resection.nii.gz'.

\subsection{Data Availability}
The proposed annotations will be freely available to public for viewing and downloading upon the acceptance of this paper via \url{https://osf.io/6y4db}. The tumor and resection cavity annotations can be freely used by research laboratories as well as clinical institutes. However, gaining any financial benefits from the distribution of the proposed annotation dataset is prohibited. The database is under the CC BY-NC-SA 4.0 License.

% #################################################### Potential Applications
\section{Discussion and Potential Applications}
The border and shape of brain tumors have long been established as  important diagnostic markers in resection surgeries. As a result, several image processing techniques have been adopted in segmentation of tumor which rely on the creation of mathematical descriptions of the tumour border. However,  validation of such techniques needs to be investigated in the case of new data. To this end, we have provided the manual annotations of brain tumor and resection cavities of iUS images from RESECT dataset. Our proposed expert-annotated dataset comprises the segmentation masks of 23 low-grade gliomas patients. We hope that publication of this expert-annotated brain iUS dataset will be beneficial to variety of research and clinical laboratories for validating and comparing deep learning algorithms as well as image registration algorithms. In this section, we explore  some of the potential applications of our proposed annotations.

The dataset's utility is primarily focused on two main applications. The first is to boost the development of novel image analysis algorithms (either based on deep learning or minimization of energy functions)  for brain tumor detection and segmentation. With an ever-growing corpus of literature presenting novel  segmentation algorithms, determining their true impact on brain images is difficult. As a result, technical and clinical communities can utilise this dataset to test their algorithms as a standard assessment metric in advancing brain cancer treatment. In addition, this dataset allows algorithms to learn how to address the task of accurate multi-instance detection and segmentation. For binary segmentation problems, there are numerous ways, however recent studies have shown that adding instances to deep learning algorithms not only it conducts the multi-task segmentation problem in parallel, but also it improves the overall performance. Consequently, we believe that the proposed dataset can further improve computer-aided diagnosis (CAD) systems for multi-instance and multi-organ analysis.

The second application is to stimulate the development, validation and benchmarking of segmentation-based registration algorithms to compensate for brain shift throughout the surgical procedure.
This will result in more accurate neuronavigation and potentially improved resection control.

To the best of our knowledge, this is the first study that publicly provides the expert-annotation segmentation of tumor and resection cavities in iUS images simultaneously. As previously explained, the time consuming and challenging procedure of delineating brain iUS images has impeded the publications of such annotations. Earlier, we discussed about several studies focusing on the use of deep learning-based algorithms on private brain US datasets. Considering the current ever-growing body of literature in applying newly proposed image analysis algorithms on brain US images, it is hard to define a standard evaluation criteria for such algorithms as their performance have remained limited in response to new US dataset. Therefore, we believe that by providing this set of iUS annotations to the public, many researchers worldwide can incorporate these annotations for setting a baseline and developing their proposed segmentation and registration algorithms. It also has the potential to greatly facilitate the growth and validation of deep learning algorithms for iUS images during operation guidance. Additionally, the presented annotations can also be used in registration methods for correcting the brain shift during the resection surgery\cite{canalini2019segmentation}.

%Moreover, as sulci remains visible during the resection procedure, its annotations can be used as a guidance landmark to ensure efficient tracking of resection surgery.

In conclusion, the proposed expert-annotated dataset presents a real-world dataset consisting of annotations of brain tumor and resection cavities in iUS images of RESECT database. The dataset can be used to both train and evaluate segmentation and registration methods, thereby challenging their ability.

% #################################################### References
\section*{References}
\addcontentsline{toc}{section}{\numberline{}References}
\vspace*{-20mm}

% Following assumes you are using bibtex. However, for submission to the
% journal you MUST explicitly INCLUDE THE REFERENCES IN THE TEX FILE.
% In that case you need the following

% \begin{thebibliography}{10}
% insert the .bbl file generated by bibtex here
	%This will be a series of entries from your .bib file formatted
	%something like
	%\bibitem{Me09}
        %{I.~Meijsing, B.~W.~Raaymakers, A.~J.~E.~Raaijmakers \it et al.},
        %\newblock {Dosimetry for the MRI accelerator: the impact of a
	%magnetic field on the response of a Farmer NE2571 ionization chamber},
        %\newblock Phys. Med. Biol. {\bf 54}, 2993 -- 3002 (2009).

% \end{thebibliography}

% The following is when using bibtex and picks up the example.bib file

%\bibliography{Explicit address of .bib file}
\bibliography{RESECT-seg}      %example.bib is on the same directory
% above points to where we find the master reference list
% and also causes the bibliography to be printed

% When creating your bibliography you should run bibtex on your local
% computer after running pdflatex on your .tex file. bibtex will
% generate a .bbl file.
% Copy the contents of this .bbl file into your main latex document,
% replacing the "\bibliography" command which was pointing at your .bib file.

% following defines style of .bbl file

%\bibliographystyle{explicit relative path to medphy.bst}
\bibliographystyle{medphy.bst}    %if this is installed on your system,
				    %it is not essential to have the    ./

% Note that you need to typeset once, then run bibtex, then typeset another
% two times to get the references working properly.

\end{document}